\documentclass[aps,reprint,superscriptaddress,nofootinbib]{revtex4-1} 
\usepackage[english]{babel}

\usepackage{amsmath, amssymb}
\usepackage{mathtools}
\usepackage{siunitx}
\usepackage{environ}
\usepackage{mathrsfs}
\usepackage{braket}
\usepackage[colorlinks]{hyperref}
\DeclareMathAlphabet{\mathscrlower}{OT1}{pzc}{m}{it} 
\usepackage{prettyref}
\newrefformat{tab}{Table \ref{#1}}
\newrefformat{fig}{Figure \ref{#1}}
\usepackage{graphicx}
\usepackage{booktabs}
\usepackage{array}
\usepackage{multirow}
\usepackage{dcolumn}
\usepackage{threeparttable}
\usepackage{threeparttablex}
\usepackage{placeins}
\usepackage[stable]{footmisc}
\usepackage{mhchem}

\newcommand{\pauli}{\boldsymbol{\sigma}}
\newcommand{\Pauli}{\boldsymbol{\sigma}}
\newcommand{\diraccontra}[1]{\boldsymbol{\gamma}^{#1}}

\newcommand{\unity}{{\bf 1}_{2\times 2}}
\newcommand{\zeroty}{{\bf 0}_{2\times 2}}
\newcommand{\pos}{\vec{r}}

\newcommand{\momop}{\hat{\vec{p}}}

\newcommand{\spinmom}{\vec{\Pauli}\cdot\momop}

\newcommand{\Sum}[2]{\sum\limits_{#1}^{#2}}

\newcommand{\parantheses}[1]{\left(#1\right)}

\let\nablatmp\nabla
\renewcommand{\nabla}{\vec{\nablatmp}}
\DeclarePairedDelimiter\abs{\lvert}{\rvert}
\makeatletter
\let\oldabs\abs
\def\abs{\@ifstar{\oldabs}{\oldabs*}}

\newcommand{\partiell}[2]{\frac{\partial #1}{\partial #2}}
\AtBeginDocument{
\heavyrulewidth=.08em
\lightrulewidth=.05em
\cmidrulewidth=.03em
\belowrulesep=.65ex
\belowbottomsep=0pt
\aboverulesep=.4ex
\abovetopsep=0pt
\cmidrulesep=\doublerulesep
\cmidrulekern=.5em
\defaultaddspace=.5em
}
\begin{document}
\title{Chiral molecules as sensitive probes for direct detection of 
$\mathcal{P}$-odd cosmic fields}
\date{\today}
\author{Konstantin Gaul}
\affiliation{Fachbereich Chemie, Philipps-Universit\"{a}t Marburg,
Hans-Meerwein-Stra\ss{}e 4, 35032 Marburg, Germany}

\author{Mikhail G. Kozlov}

\affiliation{Petersburg Nuclear Physics Institute of NRC
``Kurchatov Institute'', Gatchina 188300, Russia} 
\affiliation{St.~Petersburg Electrotechnical University ``LETI'',
Prof. Popov Str. 5, 197376 St.~Petersburg}

\author{Timur A. Isaev}

\affiliation{Petersburg Nuclear Physics Institute of NRC
``Kurchatov Institute'', Gatchina 188300, Russia} 

\author{Robert Berger}
\affiliation{Fachbereich Chemie, Philipps-Universit\"{a}t Marburg,
Hans-Meerwein-Stra\ss{}e 4, 35032 Marburg, Germany}
\begin{abstract}
Particular advantages of chiral molecules for direct detection of the
time-dependence of pseudoscalar and the timelike-component of
pseudovector cosmic fields are highlighted. Such fields are invoked in
different models for cold dark matter or in the Lorentz-invariance violating
standard model extensions and thus are
signatures of physics beyond the standard model. The sensitivity of a
twenty year old experiment with the molecule CHBrClF to pseudovector
cosmic fields as characterized by the parameter $\abs{b^\mathrm{e}_0}$
is estimated to be $\mathcal{O}(10^{-12}\,\si{\giga\electronvolt})$
and allows to predict the sensitivity of future experiments with
favorable choices of chiral molecular probes to be
$\mathcal{O}(10^{-17}\,\si{\giga\electronvolt})$, which will be an
improvement of the present best limits by at least two orders of magnitude.
\end{abstract}
\maketitle
\emph{Introduction.---}%
The nature of dark matter (DM), the existence of which is invoked to
explain the cosmological motion of visible matter, is considered to be
one of the biggest unsolved problems of modern physics (see e.g.
Ref.~\cite{bertone:2005}). Among the various DM theories, the cold DM (CDM)
variant appears to provide a simple explanation for a wealth of
astrophysical observations~\cite{davis:1985}. Up to now, however, the
constituents of CDM are unknown and can range from macroscopic objects
such as black holes to new particles like weakly interacting massive
particles (WIMPs), axions, sterile neutrinos or dark photons (see e.g.
Refs.~\cite{dodelson:1994,cheng:2002,arias:2012}). 

The model of CDM has also several
shortcomings~\cite{gentile:2004,klypin:1999,pawlowski:2014,kormendy:2010,sachdeva:2016,kroupa:2010}.
In order to overcome some of these, so-called fuzzy CDM models, which
assume  CDM to consist of ultra light particles with masses of
$m_\phi\sim\SI{1e-22}{\electronvolt}/c^2$, were
proposed~\cite{hu:2000,lee:2018}.

CDM candidates are different types of weakly interacting particles (an
overview can be found e.g. in Ref.~\cite{graham:2016}).  Among those,
we focus in the following on pseudoscalar and pseudovector particles as they are
a source of direct parity ($\mathcal{P}$) violation.

\emph{Pseudoscalar} CDM particles behave as axions, which were originally
proposed~\cite{peccei:1977,wilczek:1978,weinberg:1978} as a solution
to the so-called strong $\mathcal{CP}$-problem~\cite{hooft:1976}, i.e.
the apparently missing $\mathcal{CP}$-violation in quantum
chromodynamics (QCD) although there is a free parameter in QCD that
can account for such a violation. The window to search for such
particles can be restricted to a defined parameter space, like for the
QCD axion (see e.g.~\cite{cortona:2016}) which has to solve the
strong $\mathcal{CP}$-problem, or can be large as for axionic
particles that are not bound to solve the strong
$\mathcal{CP}$-problem. The latter are often referred to as axion-like
particles (ALPs). \emph{Pseudovector} fields are important for models such
as dark photons~\cite{an:2015,catena:2018} and also appear as sources
of local Lorentz invariance violation in the Standard Model Extension
(SME)~\cite{colladay:1998}. 

In the last decade many proposals for new experiments and improved
bounds on pseudoscalar CDM appeared, some of which employ atomic
spectroscopy (see e.g.~\cite{graham:2011,graham:2013,sikivie:2014,roberts:2014,stadnik:2014,graham:2018}).
Among the latter, direct measurement of parity violation with modern
atomic precision spectroscopy~\cite{roberts:2014,roberts:2014a}
provided strict limits on static $\mathcal{P}$-odd fields, where
effects of these fields adds to parity violating effects stemming from electroweak
electron-nucleus interactions mediated by the $Z^0$ boson.

It is well known that such $\mathcal{P}$-odd effects are strongly
enhanced in chiral molecules, as the chiral arrangement of the nuclei
leads to helicity in the electron cloud (see e.g. Refs.~\cite{berger:2004a,berger:2019}). This effect can be measured as energy difference
between enantiomers of chiral molecules or as resonance frequency
differences between the two non-identical mirror-image molecules
\cite{quack:1986,letokhov:1975}. As frequency shifts can be measured
very accurately, this appears to be a particularly promising tool to
search for $\mathcal{P}$-odd cosmic fields (for recent reviews on
molecular parity violation
see~\cite{berger:2019,schwerdtfeger:2010,quack:2008,quack:2005,crassous:2005,berger:2004a,quack:2002}).
In the following we show advantages of the use of chiral molecules to
search for $\mathcal{P}$-odd cosmic fields. We estimate the
sensitivity
on cosmic parity violation of a twenty year old experiment
\cite{daussy:1999} with the chiral methane derivative CHBrClF
\cite{kompanets:1976,bauder:1997} and discuss the prospects of modern
experiments with chiral molecules.

\emph{Theory.---}%
We write the \emph{pseudoscalar} cosmic
field as $\phi(t)=\phi_0\cos(\omega_\phi t)$ (see e.g. Ref.
\cite{roberts:2014a}), which is supposed to behave
non-relativistically
$\hbar\omega_\phi\approx m_\phi c^2$. The interaction of
electrons $\psi_\mathrm{e}$ with such pseudoscalar fields $\phi(t)$ can
be described by the following Lagrangian density (see e.g.
\cite{wilczek:1978,weinberg:1978})
\begin{align}
 \label{eq: ac_pnc1}
 \mathcal{L}^\phi_{\mathrm{ps}}
 &=
 g_{\phi\mathrm{\bar{e}e}}(\hbar c\,\partial_\mu\phi)\bar{\psi}_\mathrm{e}\diraccontra{\mu}\diraccontra{5}\psi_\mathrm{e}\,,
\end{align}
where $g_{\phi\mathrm{\bar{e}e}}$ is a coupling constant of dimension
\si{\per\giga\electronvolt}. Here the $4 \times 4$ Dirac matrices are defined 
as 
\begin{equation}
\diraccontra{0}=\begin{pmatrix}
\unity &\zeroty\\
\zeroty &-\unity\\
\end{pmatrix},\qquad
\diraccontra{k}=\begin{pmatrix}
\zeroty &\pauli^k\\\
-\pauli^k &\zeroty\\
\end{pmatrix},
\end{equation}
where $\pauli^k$ are the Pauli spin matrices with upper indices $k=1,2,3$. The
index $\mu$ runs as $\mu=0,1,2,3$. We define 
$\diraccontra{5}=\imath\diraccontra{0}\diraccontra{1}\diraccontra{2}\diraccontra{3}$
with $\imath=\sqrt{-1}$ being the imaginary unit.
$\partial_\mu=\partiell{}{x^\mu}$ is the first derivative with respect
to the four-vector $x^\mu=(ct, x, y, z)$ and we use Einstein's sum
convention here for convenience.
The time-derivative of the pseudoscalar field leads
to the $\mathcal{P}$-odd one-electron Hamiltonian 
\begin{equation}
\hat{h}_\mathrm{ps}=g_{\phi\mathrm{\bar{e}e}}
\sqrt{2(hc)^3\rho_\mathrm{CDM}}
\sin(\omega_\phi t)
\diraccontra{5},
\label{eq: alppv}
\end{equation}
where
$\rho_\mathrm{CDM}\approx\frac{(\hbar\omega_\phi\phi_0)^2}{2(hc)^3}$
is the CDM energy density, for which we assume all ALPs  to comprise all of the
CDM with a uniform density: 
$(hc)^3\rho_\mathrm{CDM}=(hc)^3~\SI{0.4}{\giga\electronvolt\per\centi\meter\cubed}=\SI{7.6e-4}{\electronvolt^4}$
(see Ref. \cite{vergados:2017}). 

Electronic interactions with \emph{pseudovector} cosmic fields
can be described by the Lagrangian density
\begin{equation}
\mathcal{L}_\mathrm{pv}^b = -b^\mathrm{e}_\mu
\bar{\psi}_\mathrm{e}\diraccontra{\mu}\diraccontra{5}\psi_\mathrm{e},
\end{equation}
which appears e.g.\ in the local Lorentz invariance violating Standard Model
Extension (SME) (for details
see Refs. \cite{colladay:1998,kostelecky:1999}).
The parity non-conserving one-electron interaction Hamiltonian for the
temporal component $\mu=0$ is
\begin{equation}
\hat{h}_\mathrm{pv}=b^\mathrm{e}_0(t)\diraccontra{5},
\label{eq: pseudovector}
\end{equation} 
where the field can be static $b^\mathrm{e}_0(t)=b^\mathrm{e}_{0}$ or dynamic
$b^\mathrm{e}_0(t)=b^\mathrm{e}_0\sin(\omega_b t)$. Here
$b^\mathrm{e}_0$ is the interaction strength of the timelike-component of
the field with the electrons.

The operators corresponding to electronic interactions with
$\mathcal{P}$-odd cosmic fields shown above are proportional to
$\diraccontra{5}$. The electronic expectation value of
$\Braket{\diraccontra{5}}$ can be expanded in orders of the fine
structure constant $\alpha$ giving in leading order:
\begin{equation}
\Braket{\diraccontra{5}}\approx\alpha\Braket{\spinmom}\,,
\label{eq: nonrelg5}
\end{equation}
where $\momop$ is the electronic linear momentum operator. 
As $\spinmom$ is
an imaginary, electron-spin dependent operator, this expectation value vanishes
in the strict electrostatic limit, but it can become non-zero when 
spin-orbit coupling $\hat{H}_\mathrm{so}$ is accounted for, similarly to
the situation for parity-violation in chiral molecules due to weak neutral 
currents \cite{zeldovich:1977,harris:1978}. Furthermore, it is
obvious from eq. \prettyref{eq: nonrelg5} that $\Braket{\diraccontra{5}}$
depends on the helicity of the electron cloud. Thus,
$\Braket{\diraccontra{5}}$ can be non-zero in a chiral molecule, in which the
electrons move in 
a $\mathcal{P}$-odd nuclear potential, whereas in a non-chiral molecule or
in an atom $\Braket{\diraccontra{5}}$ vanishes in the absence of additional
$\mathcal{P}$-odd forces. 

It can be shown from perturbation theory that for systems containing
two heavy main group elements with nuclear charge numbers $Z_A$ and $Z_B$
the following scaling relation holds in lowest order:
\begin{equation}
\Braket{\diraccontra{5}}_\mathrm{mol}
\sim c_1 \alpha^5Z_A^2Z_B^2 + c_2 \alpha^3Z_A^2 + c_3 \alpha^3Z_B^2\,,
\label{eq: scaling}
\end{equation}
Here the factor $\alpha^2Z_B^2$ in the first term emerges from
spin-orbit coupling. A detailed derivation together with evidence from
numerical studies of several chiral molecules will be provided in a
separate publication~\cite{gaul:2020d}. From this, and previous studies
of the electronic expectation value of $\diraccontra{5}$ as a possible total
molecular chirality measure \cite{senami:2019} (see, however, the
critical discussion in Ref.~\cite{ruch:1972} on the utility of
pseudoscalar functions as chirality measures), it can be deduced that
contributions at the nuclear center dominate the electronic expectation value of
$\diraccontra{5}$ and let it behave similarly to nuclear-spin
independent electroweak electron-nucleon current interactions
described by the one-electron Hamiltonian
\begin{equation}
\hat{h}_\mathrm{ew}=\frac{G_\mathrm{F}}{2\sqrt{2}}\Sum{A=1}{N_\mathrm{nuc}}Q_{\mathrm{W},A}\rho_A(\pos)\diraccontra{5}
\end{equation}
with $G_\mathrm{F}$ being Fermi's constant, $Q_{\mathrm{W},A}$ being the
weak nuclear charge of nucleus $A$ with nuclear density distribution
$\rho_A(\pos)$ and the sum running over all $N_\mathrm{nuc}$ nuclei.

Thus, molecular experiments that aim to test parity violation due to
weak interactions can also be used for searches of parity violating
cosmic fields with a comparable sensitivity.

\emph{Results and Discussion.---}%
In the following we estimate the expected sensitivity of
experiments with chiral molecules to $\mathcal{P}$-odd cosmic fields
as characterized by the $b^\mathrm{e}_{0}$ parameter
from an experiment with \ce{CHBrClF} reported by Daussy \textit{et.
al.}\cite{daussy:1999}. In this experiment the C--F stretching
fundamental vibration ($\nu_4$) in enantioenriched samples of
CHBrClF was studied by high-resolution infrared spectroscopy.  
We are interested in the parity-violating splittings of the vibrational 
resonance frequency induced by cosmic fields interacting 
through $\Braket{\diraccontra{5}}$. 

Our calculations for CHBrClF, which are described in more detail in a
separate publication~\cite{gaul:2020d}, were carried out following Ref.
\cite{berger:2007}, which utilized the separable anharmonic adiabatic
approximation framework as described in Ref.~\cite{quack:2000a}.
Parity-violating molecular properties were computed on the level of
two-component ZORA-cGKS with the B3LYP density functional.
We reuse electronic densities and Kohn-Sham orbitals as well as
vibrational wave functions determined in Ref. \cite{berger:2007}.
With these, electronic expectation values of 
$\diraccontra{5}=\sum_i\diraccontra{5}_i$ and of the
nuclear-spin independent electroweak electron-nucleon current
interaction term induced by
$\hat{H}_\mathrm{ew}=\sum_i\hat{h}_{\mathrm{ew}}(i)$ 
were calculated with our ZORA property toolbox approach outlined in
Ref.~\cite{gaul:2020}.
Vibrational corrections of the properties were computed as described
in Ref. \cite{berger:2007}.

The (negative) outcome of the experimental test for a parity violating
frequency shift reported in Ref.~\cite{daussy:1999} is
$\left|\Delta\nu\right|=\SI[separate-uncertainty=true]{9.4\pm17.9}{\hertz}$.

\begin{figure}[htp]
\includegraphics[width=.5\textwidth]{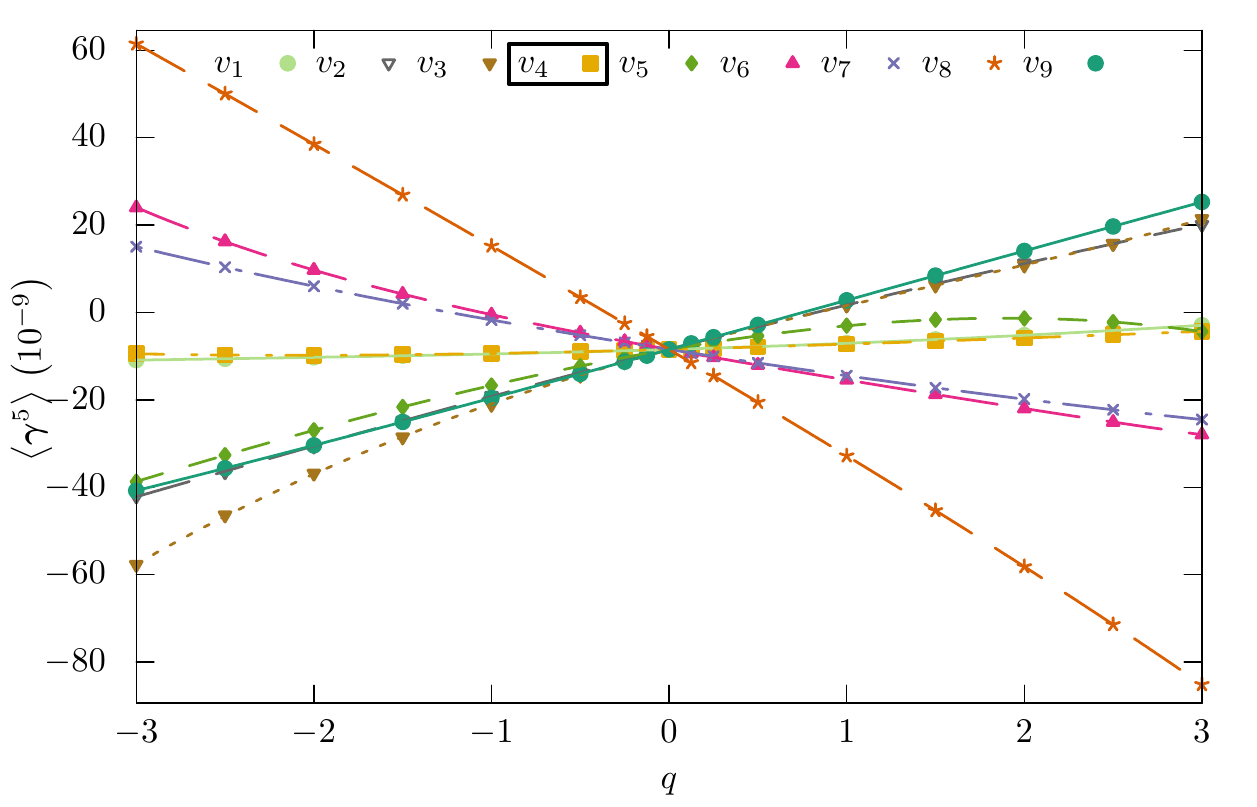}
\caption{Dependence of the expectation value of $\diraccontra{5}$ on
the nine different modes in \ce{$(S)$-CHBrClF} computed at the level
of ZORA-cGKS with the B3LYP functional and polynomial fits to
$\Braket{\diraccontra{5}}$ to fourth order (lines). The C--F
stretching mode $\nu_4$ was studied in the experiment in
Ref.~\cite{daussy:1999}.}
\label{fig: all_modes}
\end{figure}

The expectation values of $\diraccontra{5}$ for the ground and first
excited vibrational states along the C-F stretching mode of
$(S)$-CHBrClF are computed to be
\begin{align}
\Braket{v_4=0|\diraccontra{5}|v_4=0}&=-8.28\times10^{-9}\\
\Braket{v_4=1|\diraccontra{5}|v_4=1}&=-7.91\times10^{-9}\,.
\end{align}
This leads to an estimate for the splitting between the two enantiomers of 
CHBrClF due to the perturbation with $\diraccontra{5}$ for the transition 
between the vibrational ground and first excited states of $v_4$ of
\begin{multline}
\Delta_{(R,S)}\Braket{\diraccontra{5}}=2\parantheses{
\Braket{v_4=1|\diraccontra{5}|v_4=1}
\right.\\\left.
-
\Braket{v_4=0|\diraccontra{5}|v_4=0}
} \approx
7.4 \times10^{-10} .
\end{multline}
As we discuss in more detail in Ref.~\cite{gaul:2020d}
non-separable anharmonic effects can play a prominent role for the
C--F stretching mode in CHBrClF as effects characterized by the first
and second derivatives with respect to $q_4$ can be expected to be of
the same order as those characterized by first derivatives with
respect to $q_{r\not=4}$. This can best be seen from a plot of
$\Braket{\diraccontra{5}}$ on one-dimensional cuts along all modes
(see \prettyref{fig: all_modes}). Therein the weak dependence of
$\Braket{\diraccontra{5}}$ on $q_4$ in comparison to the pronounced
dependence on other modes stands out. Therefore, it is not possible
to provide a robust theoretical value for $\Braket{\diraccontra{5}}$
for the C--F stretching mode, but we give rather the order of
magnitude, which is
$\Delta_{(R,S)}\Braket{\diraccontra{5}}\sim\mathcal{O}(10^{-10})$.
The sensitivity of this experiment to $b^\mathrm{e}_{0}$ is is found
to be of the order
\begin{align}
\abs{b^\mathrm{e}_{0}}
\lesssim\left|\frac{\SI{12.7}{\hertz}}{\mathcal{O}(10^{-10})}h\right|
\sim\mathcal{O}(10^{-12}\,\si{\giga\electronvolt})
\end{align} 
This sensitivity based on the twenty year old experiment on CHBrClF is about two 
orders of magnitude inferior to the best limit from modern atomic experiments 
of \SI{7e-15}{\giga\electronvolt} so far~\cite{roberts:2014}. An improvement in
theory, most importantly by consideration of multi-mode effects
\cite{quack:2003a,gaul:2020d} and additionally by calculations with more sophisticated
electronic structure methods, would allow to place a robust limit as
we have highlighted in Ref.~\cite{gaul:2020d}. 

The sensitivity of the molecular experiment is supposed
to be improvable by two orders of
magnitude or better by a different experimental setup as discussed in
Refs.~\cite{ziskind:2002,darquie:2010,cournol:2019}. The scaling behaviour in \prettyref{eq:
scaling} suggests that further sensitivity improvements are possible by
selecting heavy-elemental chiral molecules. Electroweak $\mathcal{P}$-odd
effects, which scale like $N_A Z_A^2 Z_B^2$ with $N_A$ being the number of
neutrons of nucleus $A$, were estimated to give vibrational splittings that
can become three orders of magnitude larger in well-chosen heavy-elemental
molecules when compared to CHBrClF \cite{berger:2007,darquie:2010}. Due to
the missing $N_A$ scaling, an enhancement by two orders of magnitude can
thus be anticipated for $\Delta_{(R,S)}\Braket{\diraccontra{5}}$. 
Furthermore, as indicated in \prettyref{fig: all_modes} and highlighted in
Ref.~\cite{gaul:2020d}, the sensitivity is improvable by an order of
magnitude by choice of a different vibrational transition. 

Thus we can estimate that in future $\mathcal{P}$-violation
experiments with chiral molecules the sensitivity of the 1999 experiment
can be improved by at least five orders of magnitude down to
$10^{-17}\,\si{\giga\electronvolt}$, i.e.\ an improvement of the actual
best limit by at least two orders of magnitude. This renders
experiments with suitably chosen chiral molecules sensitive probes for
physics beyond the Standard Model.

To exploit its full potential, however, a measurement of cosmic parity
violation on the background of the larger electroweak frequency
splittings would become necessary, which makes additional demands on
accuracy of the accompanying computational approaches or calls for
experimental schemes to disentangle these two contributions for
instance by measuring isotope-dependent electroweak frequency
splittings.

The experiment discussed above is sensitive to \emph{oscillating}
$\mathcal{P}$-odd interactions of electrons as well.  We can exploit
the fact that the experiment was performed over a time span of ten
days with a well defined set of measurements on each day.  In the
following we estimate expected sensitivities for this kind of
experiments to oscillating pseudoscalar and pseudovector fields. As
CHBrClF is not an optimal choice, we do not aim to determine the best
possible limit from the actual experiment but rather highlight the
applicability of such a type of experiment for the direct detection of
oscillating pseudovector fields.

The measured frequency shift due
to electronic interactions with ALP fields is proportional to
\begin{equation}
g_{\phi\mathrm{\bar{e}e}}
\sqrt{2(hc)^3\rho_\mathrm{CDM}}
\sim
\SI{4e-20}{\giga\electronvolt\squared}g_{\phi\mathrm{\bar{e}e}}
\end{equation}
For \emph{pseudoscalar} fields, measurements of the time-derivative of
the ALP field as well as the spatial-derivatives are sensitive to the
same parameter $g_{\phi\mathrm{\bar{e}e}}$. Thus, it would require
static bounds on the order of $10^{-30}\,\si{\giga\electronvolt}$
(i.e. a precision of \si{\micro\hertz} in the CHBrClF experiment) to
be competitive with spin precession experiments that set limits of
$\abs{g_{\phi\mathrm{\bar{e}e}}}<10^{-7}\si{\per\giga\electronvolt}$
(see Refs.  \cite{graham:2018,bloch:2019}). This appears not to be
achievable with experiments available today that follow this approach
for chiral molecules.

Chiral molecules, however, are directly sensitive to the
timelike-component of oscillating \emph{pseudovector} fields, which is
not favorably accessible in spin precession experiments. In the
following we discuss briefly the expected sensitivity on
$b_0^\mathrm{e}$ of oscillating fields that can in principle be
obtained from available experiments with chiral molecules.

To obtain a rough estimate for the sensitivity to $b^\mathrm{e}_0$ in
dependence of $\omega_b$ due to the sinusoidal behavior of
$b_{0}^\mathrm{e}(t)$ we assume that the sensitivity is decreasing for
larger frequencies with $\omega_b t_\mathrm{tot}$. Furthermore we can
expect that the experimental uncertainty increases with resulting
shorter interrogation times for larger $\omega_b$ as
$\sim\sqrt{\omega_b t_\mathrm{tot}}$ and we expect the experiment not
to be sensitive to frequencies with $\omega_b
t_\mathrm{tot}>n_\mathrm{tot}$, where $n_\mathrm{tot}$ is the total
number of individual measurements. As CDM is supposed to be incoherent
for small frequencies $\omega_b < 2\pi/t_\mathrm{tot}$ we can expect
that $b^\mathrm{e}_0$ converges to the static limit. The experiment in
Ref.~\cite{daussy:1999}
was performed on 10 separate days with a total of 580 individual
measurements. When assuming a continuous measurement campaign on each
day of 58 subsequent measurements we have
$t_\mathrm{tot}\approx\SI{1}{\day}$ and $n_\mathrm{tot}\approx58$. In
total we arrive at the sensitivities 
\begin{equation}
b^\mathrm{e}_0\lesssim\begin{cases}
10^{-12}\,\si{\giga\electronvolt}, &\mathrm{if}~\frac{\omega_b}{2\pi} \leq
\SI{1.2}{\micro\hertz}\\
(\omega_b t_\mathrm{tot})^{3/2}10^{-12}\,\si{\giga\electronvolt}, &
\mathrm{if}~\SI{1.2}{\micro\hertz}<\frac{\omega_b}{2\pi}\leq
\SI{0.7}{\milli\hertz}\\
\infty, & \mathrm{if}~\frac{\omega_b}{2\pi} >
\SI{0.7}{\milli\hertz}
\end{cases}
\,.
\end{equation}

\begin{figure}
\includegraphics[width=.5\textwidth]{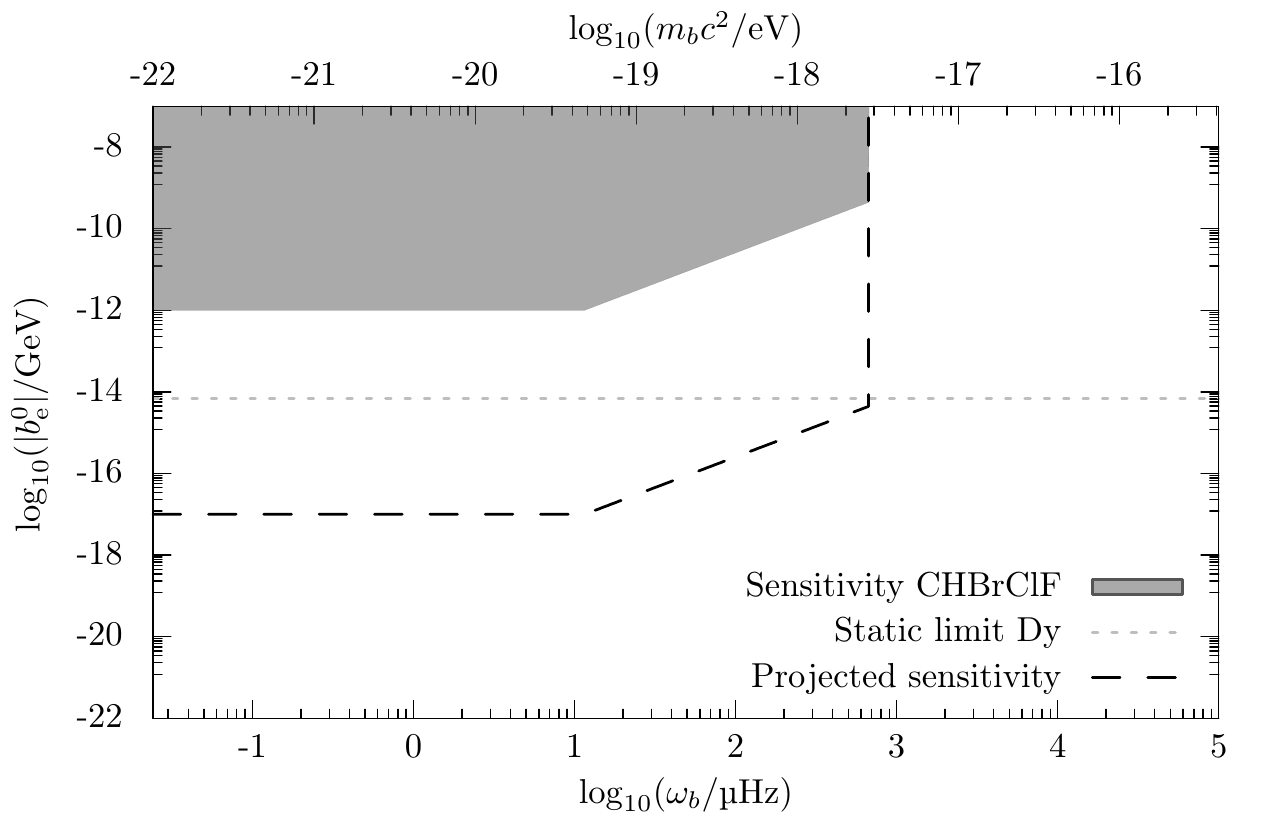}
\caption{
Sensitivity on electron couplings with the timelike-component of
pseudovector fields $b_0^\mathrm{e}$ in dependence of the CDM
pseudovector oscillation frequency $\omega_b$ from a twenty year old
experiment with CHBrClF\cite{daussy:1999} compared to the actual best
static limit on $b_0^\mathrm{e}$ from the Dy experiment (see Ref.
\cite{roberts:2014}). The projected sensitivity is achievable with modern
experiments with chiral molecules, assuming an improvement in
sensitivity of 5 orders
of magnitude compared to the CHBrClF experiment of 1999.
}
\label{fig: dynlimit}
\end{figure}

The expected sensitivities on $b^\mathrm{e}_0$ in CHBrClF and future
experiments in dependence on the pseudovector CDM oscillation
frequency $\omega_b$ is shown in \prettyref{fig: dynlimit}. It shall
be noted that the region of $\omega_b$ to which the experiment is
sensitive may be smaller or even extended depending on the actual
timing of the measurements. However, robust bounds require an 
extended theoretical description and a rigorous statistical analysis 
of the actual data sets as was also discussed in Refs.
\cite{adelberger:2019,wu:2019,centers:2019}. 

\emph{Conclusion.---}%
We have shown in this letter that $\mathcal{P}$-odd interactions of
electrons with cosmic fields are strongly pronounced in chiral
molecules.  We could demonstrate that chiral molecules are suitable
systems to tighten bounds on $\mathcal{P}$-odd electronic interactions
of static pseudovector fields that emerge e.g.\ from the Standard Model
Extension. By performing quasi-relativistic calculations of
expectation values of $\mathcal{P}$-odd cosmic field interactions in
CHBrClF including vibrational corrections, we demonstrated that the
C--F stretching mode is not a good choice to place robust limits on
$\mathcal{P}$-odd cosmic fields as the effects are
comparatively small and also difficult to prediction due to pronounced
multimode contributions. However, we estimated the sensitivity
of this mode to the parameter $b^\mathrm{e}_0$ to be on the
order of $10^{-12}\,\si{\giga\electronvolt}$ in a 20 year old
experiment. This sensitivity is inferior by two orders of magnitude to
the actual best direct measurements drawn from modern atomic parity
violation experiments. We estimate the achievable sensitivity to
$\mathcal{P}$-odd cosmic fields with modern high-resolution molecular
spectroscopy on suitably chosen chiral molecules to be on the order of
$10^{-17}\,\si{\giga\electronvolt}$ for static fields (see
\prettyref{fig: dynlimit}). This would be an improvement of the
current best limit on $b^\mathrm{e}_0$ by two orders of magnitude.
Furthermore, we discussed possibilities of direct detection of ultra
light dark matter by studying oscillating parity violating potentials
in chiral molecules. We have shown that without design of a
fundamentally new experimental concept limits on electronic
interactions of ultra light oscillating pseudovector particles
$b^\mathrm{e}_{0}$ with frequencies of around
$\omega_b\lesssim\SI{10}{\micro\hertz}$ could be pushed to about
$10^{-17}\,\si{\giga\electronvolt}$ or better with modern experiments
with chiral molecules. This corresponds to a direct detection of CDM
masses below $10^{-19}\,\si{\electronvolt}/c^2$ and thus can be
interesting for fuzzy dark matter searches.

\begin{acknowledgments}
The authors are grateful to the Mainz Institute for Theoretical
Physics (MITP) for its hospitality and its partial support during the
completion of this work. The Marburg team gratefully acknowledges computer
time provided by the center for scientific computing (CSC) Frankfurt
and financial support by the Deutsche Forschungsgemeinschaft via
Sonderforschungsbereich 1319 (ELCH) ``Extreme Light for Sensing and
Driving Molecular Chirality''. The work of M.G.K. and T.A.I. was supported by the
Russian Science Foundation (RSF) grant No. 18-12-00227. R.B. acknowledges
discussions with Nils Huntemann on tests of Lorentz symmetry.
\end{acknowledgments}


%
\clearpage


\end{document}